\documentclass[
doubleblind]{ceurart}

\usepackage{listings}
\usepackage{enumitem}
\usepackage{multirow}
\usepackage{amsmath}
\usepackage{caption}
\usepackage{graphicx}
\usepackage{subcaption}
\usepackage{arydshln}

\begin{document}


\copyrightyear{2024}
\copyrightclause{Copyright for this paper by its authors.
  Use permitted under Creative Commons License Attribution 4.0 International (CC BY 4.0).
CEUR Workshop Proceedings (CEUR-WS.org)}

\conference{MuRS 2024: 2nd Music Recommender Systems Workshop, October 14th, 2024, Bari, Italy}

\title{Navigating Discoverability in the Digital Era: A Theoretical Framework}



\author[1]{Rebecca Salganik}[orcid=0009-0007-9273-8780,email=rsalgani@ur.rochester.edu,
]
\address[1]{University of Rochester, Rochester, NY, USA}

\author[2]{Valdy Wiratama}[orcid=0009-0006-6389-2335,
email=Valdy.Wiratama@vub.be,
]
\address[2]{imec-SMIT, Vrije Universiteit Brussel, Pleinlaan 9 1050 Brussels, Belgium}

\author[2]{Adelaida Afilipoaie}[orcid=0000-0002-8866-8536,
email=Adelaida.Afilipoaie@vub.be, 
]
\author[2]{Heritiana Ranaivoson}[orcid=0000-0002-3909-0440,
email=Heritiana.Renaud.Ranaivoson@vub.be,
]


\begin{abstract}
The proliferation of digital technologies in the distribution of digital content has prompted concerns about the effects on cultural diversity in the digital era. The concept of discoverability has been presented as a theoretical tool through which to consider the likelihood that content will be interacted with. The multifaceted nature of this broad theme has been explored through a variety of domains that explore the ripple effects of platformization, each with its own unique lexicography. However, there is yet to be a unified framework through which to consider the complex pathways of discovery. In this work we present the discovery ecosystem, consisting of six individual, interconnected components, that encompass the pathway of discovery from start to finish. 
\end{abstract}



\maketitle

\vspace{-.2in}
\section{Introduction}
\vspace{-.1in}
As our interactions with creative multimedia content shift towards online platforms, the algorithmic systems that guide our interactions with various art forms play an increasingly influential role in shaping how content is created, communicated, and consumed \cite{baym_making_2021}. The seemingly endless amount of content has prompted the integration of algorithmic curation \cite{villermet_follow_2021} into almost every media platform, while the effects of this algorithmic curation have impacted various media domains, including audiovisual \cite{ranaivoson_online_2019}, musical \cite{hesmondhalgh_impact_2023}, journalistic \cite{shin_countering_2022, vrikenhoek_diversity_2024}, or otherwise. Fundamentally, the goal of these digital technologies is to aid
users in their exploration of extensive content catalogs. On the one hand, algorithmically-driven “platformization” has promised to create opportunities for users to consume diverse content which, in turn, could broaden their engagement horizons. However, in actuality, various analyses have raised concerns that continued interaction with algorithmic curation can diminish diversity in the cultural expressions \cite{celma_music_2010}, creative aesthetics \cite{bernstein_diversity_2021}, and the creator demographics from whom content comes \cite{ferraro_break_2021}. While content providers may promise their users a sense of boundless abundance, many researchers have concluded that the reality in which users engage with these algorithmic systems is much murkier \cite{ferraro_break_2021, abdollahpouri_addressing_2020, bauer_music_2017, seaver_algorithms_2022}. Thus, these concerning patterns have raised questions surrounding the mechanisms which control how content is discovered. 

Furthermore, the question of discovery itself is also deeply complex. Put simply, the discoverability of an item can be defined as \textbf{the likelihood that content will be interacted with}. A discovery occurs when a user interacts with an item, ideally in a way that brings them positive satisfaction \cite{garcia-gathright_understanding_2018}. Depending on the setting this item can be new (an initial discovery) or previously encountered (in the case of a re-discovery) \cite{mok_dynamics_2022, sguerra_discovery_2022}. And, there are many different lenses through which discovery can be discussed, including policy-oriented \cite{guevremont_mesures_2019, ministere_de_la_culture_de_france_mission_2020, european_commission_directorate_general_for_research_and_innovation_european_2022}, socio-technical \cite{freeman_dont_2022, ranaivoson_platforms_2023, freeman_personalised_2023}, ethnographic \cite{seaver_algorithms_2022}, and technical \cite{abdollahpouri_addressing_2020, garcia-gathright_understanding_2018, sguerra_discovery_2022} among others, with each having its own unique vocabulary for engaging with the underlying topics that are arranged under this broad concept. The purpose of this framework is to survey these discussions and highlight key contributions such that they may better serve future policy initiatives. To better illustrate the theoretical implications of this framework, we present the \textit{Discovery Ecosystem}, which traces the pathway of online discovery in  the music sector. This choice is motivated by the sector's relative abundance of literature as compared with other media formats and distribution logics. Additionally, while the scope of this work is centered around the music domain, we posit that our framework is translatable to other sectors such as books \cite{murray_secret_2021}, audiovisual \cite{ranaivoson_assessing_2010}, and many more.

\vspace{-.2in}
\section{Relevant Works}
\vspace{-.1in}
\textbf{Exposure}
One of the primary features of discovery is the visibility which is allocated to various content mediums. An important field of research encompassing both technical and sociotechnical domains revolves around the concept of \textit{exposure} where the feedback loop between the content that is served to users and the content that they end up engaging with is heavily analyzed through a variety of disciplines \cite{ranaivoson_platforms_2023, diaz_evaluating_2020, towse_cultural_2020, mazzoli_prioritisation_2020}. This notion is implicitly tied to that of discoverability because a user's engagement with creative content is mediated by their access to and awareness of it \cite{napoli_exposure_2011}. \\  
\textbf{Diversity}
An important facet of the discourse surrounding discoverability is the analysis of which content is ultimately engaged with. One of the important perspectives through which discoverability is assessed is that of diversity, both in the content that is created \cite{mazzoli_online_2020} and that which is consumed \cite{ranaivoson_assessing_2010}. Particularly within the policy domain, there have been significant efforts to understand the impacts of digital technologies on cultural diversity \cite{octavio2016impact, pasikowska-schnass_digital_2021}. Many works within the technical domain have also worked to define metrics for diversity that can either be used to train an algorithmic system or evaluate its performance \cite{schedl_tailoring_2015, ge_beyond_2010,vargas_coverage_2014, baracskay_diversity_2022}.\\ 
\textbf{Fairness}
In relation to the previously mentioned topics of exposure and diversity, the machine learning community has framed the topic of discovery within discussions surrounding the presence of algorithmic biases and methodologies for resolving the issues by designing more holistic objectives for training algorithmic systems \cite{ge_beyond_2010, Chen_exploration_2021}. In particular, many works have explored the effects of popularity bias on the shrinking diversity of content that is served to users \cite{salganik_fairness_2024, abdollahpouri_addressing_2020, celma_hits_2008, diaz_evaluating_2020}. While these works are often not explicitly addressing the concepts of discovery, they are intrinsically tied based on the assumption that in order for the discovery process to begin, there must exist the possibility of exposure. \\ 
\textbf{Exploration}
Another dimension of research within the technical community has addressed the exploration practices of users as they engage with a cataloged. These works have been framed through the lens of two, intersecting fields: visualizations \cite{petridis_tastepaths_2022, liang_interactive_2021} or the design of agentic systems that iteratively feed content to users \cite{marketplace, Chen_exploration_2021}. The human-computer interaction and user interface communities have focused on understanding the various mediums which can be used to present content. Meanwhile, the reinforcement learning community has arranged the discussion of discovery within a broader paradigm of the field that addresses the trade-offs between exploration (i.e. introducing a user to new content) versus exploitation (i.e. capitalizing on the content a user has already explicitly found relevant). \\ 
\textbf{Discovery} McKelvey and Hunt \cite{mckelvey_discoverability_2019} present a theoretical framework containing surrounds, or interfaces through which users are presented with content, vectors, or mechanisms which coordinate which content is presented, and experiences. However, their work does not account for the broader cultural consequences of the choices made by recommendation systems and the engineers who construct them. In his book Nick Seaver  presents an ethnographic study of music engineer’s perspectives on music discovery, but ultimately does not present a cohesive lens through which to make policy evaluations \cite{seaver_algorithms_2022}. Finally, Gathright et al. \cite{garcia-gathright_understanding_2018} perform an empirical user study to understand the different kinds of user groups on Spotify and their expectations for discovery and curatorial practices. 
\vspace{-.2in}
\section{Framework}
\vspace{-.1in}
\begin{figure}
     \centering
\includegraphics[width=0.99\linewidth]{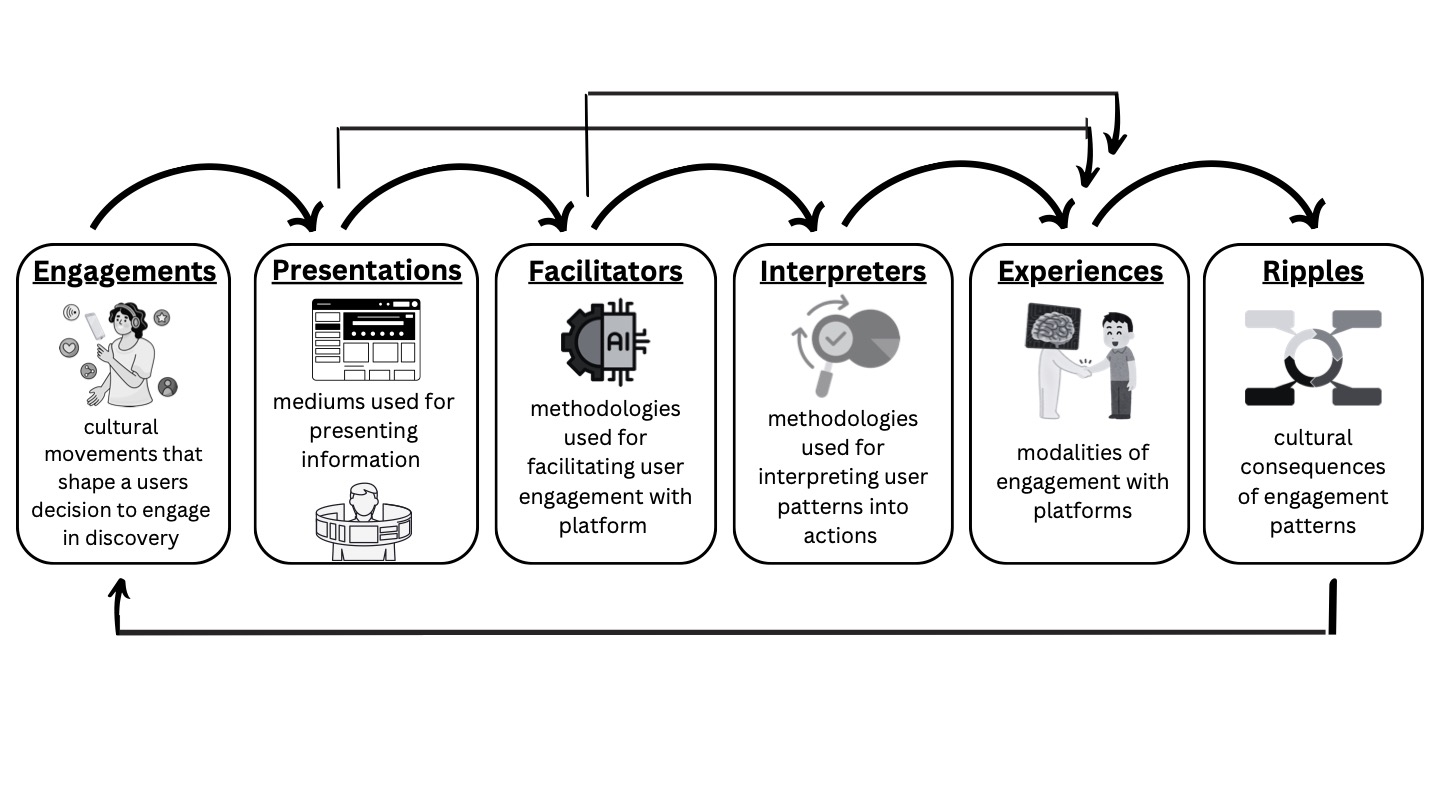}
     \vspace{-.4in}
     \caption{Visualization of Discovery Ecosystem}
        \label{fig:framework}
        \vspace{-.3in}
\end{figure}

In this section we present the six components of our framework. To ensure that all of the variables that encompass discoverability are addressed, we present six components which form what we label as the \textit{Discovery Ecosystem}. Firstly, Engagements (1) refers to the institutional activities that impact user decisions in discovery. Next, we unravel the interrelated elements of curatorial practices to define Presentations (2), Facilitators (3), Interpreters (4), as well as Experiences (5). Together, these elements encompass the process for selecting content (Facilitators), visualizing this content (Presentations), interpreting engagement patterns with said content (Interpreters), and the summation of this experience (Experiences). Finally, Ripples (6) addresses the downstream consequences that shape discovery movements within the broader online space. The purpose of organizing the framework around these six components is to present a holistic vocabulary that spans several disciplines. Engagements and Ripples encompass the large array of work that has been done in the sociotechnical and policy-related fields to understand both the motivators and impacts that are associated with the process of discovery. Meanwhile, Presentations, Facilitators, and Interpreters encompass the more technical fields of human-computer interaction, user experience, computer science, and data analysis which focus primarily on prototyping and deploying the systems that facilitate digital discovery experiences. Finally, experiences is oriented around ethnographic and psychology-oriented research into the relationships that are formed between users and the digital environments in which they engage with content discovery. As we can see from Figure \ref{fig:framework}, each of these components is heavily interrelated to the rest of the ecosystem in which it is situated. Finally, we acknowledge that there is much room for such an ecosystem to become infinitely more granular. Given the novel nature of our work, we propose these six components as a starting point and leave deeper distinctions and further analysis for future work. 


\vspace{-.1in}
\subsection{Engagements}
\vspace{-.1in}
This component encompasses the institutional activities that shape a user’s decision to engage in discovery. Before users can embark on a journey of discovery, a variety of forces shape their desire, ability, and opportunity to do so. These factors exist at multiple scales, which are not fully independent of one another. On an individual level, demographic axes, such as socioeconomic status \cite{de_vries_what_2022}, age,  community size of residence \cite{lee_understanding_2023},  neurodiversity \cite{dechaene_virtual_2023}, and technological literacy \cite{johnson_audience_2023}, are the primary method through which platforms determine a user's ability to engage in discovery within online spaces. Alternatively, on a broader scale, various musical practices can lend themselves more or less easily to digitization, thus affect their participation in discovery. While Western music has been largely translated to digital platforms, other genres have resisted their adaptation into online spaces \cite{european_expert_network_on_culture_and_audiovisual_eenca_platforms_2019}.  Finally, institutional policies can powerfully shape what kinds of practices are prioritized and available for discovery \cite{council_of_europe_committee_of_ministers_recommendation_2017}. Here, policies can shape public support for digitization \cite{european_commission_open_2020}, best practices for digitization \cite{interreg_europe_policy_2018}, and platform development \cite{european_commission_report_2018}.


\vspace{-.1in}
\subsection{Presentations}
\vspace{-.1in}
This component encompasses the mediums used to present information, focusing on various design choices for organizing content within online interfaces. It includes decisions from font and color to overall feel and navigation signals provided to users \cite{knees_intelligent_2019}. Essentially, presentations can be considered as “all that greets a user” at the face of a website or app, including content organization and navigational options \cite{morris_control_2015, besseny_lost_2020}. Corresponding to this are the spacing, sizing, placement, and relational positions of elements on a screen that are made up of intentional choices and influence the consumption of content. 
Crucially, we can see that how information is presented affects the likelihood of interaction with it. For example, in the music domain, the effect of song placement on established viewership within playlists is noteworthy \cite{maaso_streaming_2022, descottes_revenue_2022, aguiar_platforms_2021}. This, in turn, creates a feedback loop in the kind of content that will be prioritized in the future. Moreover, Ferraro et al. \cite{ferraro_break_2021} simulate a recommender system that combats bias related to gender representation in music recommendations and showcasing how several iterations of engagement with a debiased recommender system can boost representation of female artists among a broad listener group. Since the ultimate goal of interfaces is to keep users “fixed in transmission” \cite{mckelvey_discoverability_2019}, the dynamic and reactive nature of modern interfaces serves as the primary touchstone for the machinations of prioritization that are happening under the hood of a discovery experience \cite{morris_control_2015}. 

\vspace{-.1in}
\subsection{Facilitators}
\vspace{-.1in}
This component encompasses the methods through which content is curated on platforms. We intentionally select the term Facilitator to encompass both algorithmic and human curation practices. Crucially, the purpose of this paper is to understand the survey these different facilitators and contextualize their role in the discovery ecosystem, rather than providing the precise configuration of any algorithmic system.  Currently, most of the Facilitators, which are actually deployed on streaming platforms, exist in a grey area that lies at the intersection between these human and algorithmic actors \cite{bonini_first_2019, born_artificial_2021}. This is because many streaming platforms will employ human curators to guide the tastes of their algorithmically generated content \cite{pandora}. While the most prominent forms of algorithmic systems that participate in the discovery pipeline are recommender systems, more recently, the proliferation of generative conversational models, such as ChatGPT, have also taken on the responsibility of facilitating engagement with cultural content \cite{noauthor_how_2023}. Within each of the systems that a user can engage with, there is a web of different forms of algorithms that collaborate to shape their interactions. This includes retrieving information via search engine, filtering out unwanted material, producing content, and providing lists of potential options for consumption via recommendation algorithms \cite{hesmondhalgh_impact_2023}. Ultimately, these systems translate aesthetic tastes and values into computer-interpretable language, which is a key element in the discovery process. It forces certain requirements on the formats of the online content, such as music, that are being presented to users. Furthermore, this practice has been termed \textit{datafication}, and it has raised concerns throughout the various art and cultural domains \cite{born_artificial_2021, pedersen_datafication_2020, raff_music_2020, prey_nothing_2018, besseny_lost_2020}. 


\vspace{-.1in}
\subsection{Interpreters}
\vspace{-.1in}
This component encompasses the various mechanisms which are used to analyze and interpret user-content interactions within an online platform. As mentioned in the discussion related to the previous component, Facilitators, the process of content curation, both algorithmic and human, relies heavily on predicting what users will engage with. In the context of music streaming, engagement is often implied via the skips, likes, clicks, and playlists that users build. These actions are then extrapolated to interpret, shape, and facilitate a user's listening and predict their musical tastes \cite{pedersen_datafication_2020, chodos_what_2019}. It forms a cycle, a big component of which is the data standardization requirements presented within the Facilitators component. Furthermore, the reason behind this is because the metrics used to evaluate user engagement rely on incomplete information and there are many underlying assumptions embedded in the interpretation of user consumption patterns. For example, if a user skips a song, this may not necessarily imply that it is not aligned with their musical taste \cite{pedersen_datafication_2020}. Moreover, distilling something as complex as musical taste into a binary choice of liking and disliking can nudge algorithms away from helping users' exploratory patterns \cite{pedersen_datafication_2020}. Meanwhile, the complex reliance of various stakeholders on the availability and interpretability of engagement metrics has created a standardization effect, often referred to as ‘datafication’, in which everything must be represented in a form that is interpretable to an algorithmic. For example, Robert Prey highlights that user categorization on streaming platforms is shaped by advertiser and brand demands \cite{prey_nothing_2018}. In this way, a feedback loop is created between the content which is served (Facilitators), the ways in which this content is presented (Presentations), and the engagement patterns that ensue (Interpreters).

\vspace{-.1in}
\subsection{Experiences}
\vspace{-.1in}
This component encompasses modalities through which users experience their engagement with platforms. The content of this component represents the sum total of all the intended user impacts that the previous definitions provoke and thus, is intrinsically tied to Presentations, Facilitators, and Interpreters because of Experiences. Concretely, this component is focused on understanding the factors which influence a user’s perception of their own engagement in the activity of discovery. Several works have explored the different personas that users present when discovering music \cite{trocchia_typology_2011, mok_dynamics_2022, garcia-gathright_understanding_2018,lee_understanding_2015, freeman_personalised_2023}. Specifically in the music domain, many researchers have focused specifically on codifying the level of engagement that a user presents in relation to the curatorial algorithms available to them \cite{mok_dynamics_2022, garcia-gathright_understanding_2018, drott_music_2018}, contrasting between active and passive listening. However, it is important to understand that each individual discovery session can be independently understood. The same user can exhibit different patterns, depending on the context in which they are engaging with platform, and the same user can express different levels of openness towards the recommendations of a curatorial system, depending on their expectations. Furthermore, user experiences are shaped by their relationships with Facilitators which, in turn, affects their attitudes towards discovery. Researchers have shown that users form relationships with algorithmic systems, expressing interactions in relation to uman-like factors such as trust, betrayal, and intimacy \cite{freeman_dont_2022}.

\vspace{-.1in}
\subsection{Ripples}
\vspace{-.1in}
This component encompasses the cultural consequences of engagement patterns, where discoverability, and the patterns associated with it, has potential consequences. These consequences resonate not just at the individual level, but also on cultural products, the artists and producers who create them, surrounding communities, regions, and nations. When there is an improvement in discoverability, the improvement itself has generally been linked to increases in music consumption \cite{aguiar_let_2017}, but it also impacts diversity in multiple dimensions. This includes the diversity of songs \cite{datta_changing_2017}, artists \cite{aguiar_let_2017}, languages \cite{coalition_for_the_diversity_of_cultural_expressions_challenge_2020}, ideas and perspectives \cite{helberger_challenging_2018}, and neurodiversity, particularly through the incorporation of new virtual reality interfaces \cite{dechaene_virtual_2023}. Moreover, discoverability can shape the self-image of both users and cultural producers. While this component is an area that is less well-researched, a study by Robert Prey suggests that music discoverability has the potential to shape the self-image of consumers \cite{prey_nothing_2018}. The metrics derived from user engagement patterns can also influence the self-image of artists, while the invisibility with which platforms define individuals carries significant consequences. Whether we are music makers or listeners, what we discover has the power to shape how we see ourselves as individuals \cite{prey_performing_2020}. This assertion suggests that there are ripples that cannot be easily quantified in terms of pluses or minuses, as they also qualitatively affect the involved parties.
\vspace{-.2in}
\section{Conclusion}
\vspace{-.1in}

In this work we introduced the notion of a \textit{Discovery Ecosystem}, designing a framework for understanding the various components that participate in online discovery of cultural content. The purpose of this work is to survey the various disciplines that have provided perspectives on the notion of discovery, and align them within a single trajectory. This work addresses a pressing need for theoretical frameworks that are detailed enough to facilitate the design of policy initiatives for the improvement of discovery practices and the enhancement of the discoverability of content. Indeed, in addition to filling a conceptual gap, the framework can be used in policy and business contexts. From a policy perspective, the objective of discoverability is becoming more and more important in national and international cultural policy discussions, stemming from its rise to prominence in the context of Canadian markets \cite{ministere_de_la_culture_de_france_mission_2020}. Furthermore, discoverability has, in recent years, also become a business objective for companies occupying the music streaming domain whose focus on attracting users to platforms has raised the importance of offering diverse content \cite{raff_music_2020, drott_why_2018}. Thus, our holistic approach to the \textit{Discovery Ecosystem} aims to facilitate the design of policies that can improve access to and consumption of diverse content. Ultimately, this framework is intended to provide a basis for scholars from diverse disciplines to engage on the topic of discovery by using the common vocabulary provided by our six components.

\begin{acknowledgments} Research for this piece has been produced under the contract “EAC/2023/OP/0004 - Discoverability of Diverse European Cultural Content in the Digital Environment” with the European Union. The opinions expressed are those of the authors only and do not represent the contracting authority’s official position.
\end{acknowledgments}

\bibliography{sample-ceur}




\end{document}